\documentclass[%
aps, physrev,
amsmath,amssymb,
reprint,
]{revtex4-2}

\usepackage{graphicx}  
\usepackage{dcolumn}   
\usepackage{bm}        
\usepackage{siunitx}   
\usepackage{hyperref}  
\usepackage{orcidlink} 
\usepackage{aas_macros}

\DeclareSIUnit{\parsec}{pc}
\DeclareSIUnit{\year}{yr}
\DeclareSIUnit{\Msol}{\ensuremath{\mathrm{M}_\odot}}
\DeclareSIUnit{\h}{\mathit{h}}
\begin{document}

\preprint{APS/123-QED}

\title{\textbf
{\textit{Fermi}-LAT Galactic Center Excess morphology of dark matter\\in simulations of the Milky Way galaxy}
}%

\author{Moorits Mihkel Muru\orcidlink{0000-0002-4307-6094}}
  \email[Contact author: ]{mmuru@aip.de}
  \affiliation{Leibniz-Institut für Astrophysik Potsdam (AIP), An der Sternwarte 16, D-14482 Potsdam, Germany}
  \affiliation{Tartu Observatory, University of Tartu, Observatooriumi 1, 61602 Tõravere, Estonia}
\author{Joseph Silk\orcidlink{0000-0002-1566-8148}}
  \affiliation{Institut d’Astrophysique de Paris, UMR 7095, 98 bis Boulevard Arago, F-75014 Paris, France}
  \affiliation{Department of Physics and Astronomy, The Johns Hopkins University, Baltimore, MD, USA}
  \affiliation{Sub-department of Astrophysics, University of Oxford, Keble Road, Oxford OX1 3RH, UK}
\author{Noam I. Libeskind\orcidlink{0000-0002-6406-0016}}
  \affiliation{Leibniz-Institut für Astrophysik Potsdam (AIP), An der Sternwarte 16, D-14482 Potsdam, Germany}
\author{Stefan Gottlöber\orcidlink{0000-0003-4667-3174}}
  \affiliation{Leibniz-Institut für Astrophysik Potsdam (AIP), An der Sternwarte 16, D-14482 Potsdam, Germany}
\author{Yehuda Hoffman\orcidlink{0000-0002-8158-0566}}
  \affiliation{Racah Institute of Physics, Hebrew University, Jerusalem 91904, Israel}

\date{\today}

\begin{abstract}
The strongest experimental evidence for dark matter is the Galactic Center gamma-ray excess observed by the \textit{Fermi} telescope and even predicted prior to discovery as a potential dark matter signature via WIMP dark matter self-annihilations. However, an equally compelling explanation of the excess gamma-ray flux appeals to a population of old millisecond pulsars that also accounts for the observed boxy morphology inferred from the bulge old star population. We employ a set of Milky Way-like galaxies found in the \textsc{Hestia} constrained simulations of the local universe to explore the rich morphology of the central dark matter distribution, motivated by the GAIA discovery of a vigorous early merging history of the Milky Way galaxy. We predict a significantly non-spherical gamma-ray morphology from the WIMP interpretation. Future experiments, such as the Cherenkov Telescope Array, that extend to higher energies, should distinguish between the competing interpretations.
\end{abstract}

\maketitle


\section{Introduction}
The leading model for dark matter (DM) appeals to weakly interacting massive particles (WIMPs) that were initially motivated as being the only stable relic of supersymmetry (SUSY) (see the recent review by \cite{Cirelli:2024ssz}). These Majorana particles were initially in thermal equilibrium in the early Universe while relativistic, but their abundance relative to that of the radiation (photons and neutrinos) froze out once the radiation temperature fell below $kT\sim 0.1 m_\chi c^{2}$. They continued to self-annihilate, however, at a low rate until the present epoch. The predicted cross-section from the theory of weak interactions is $\sigma \propto m_\chi^{-2}$. Required freeze-out at the observed DM density $\Omega_{DM} = 0.26$ \cite{Planck:2018} predicts the annihilation cross-section. That turns out to be what is generically expected by quantum electrodynamics and is often dubbed the WIMP miracle. Integrating over the velocity dispersion of halo DM particles, one obtains the thermal freeze-out cross-section for annihilation into various standard model high-energy particles and photons. The WIMP hypothesis was recognized to predict annihilation channels whose high-energy signals are potentially detectable. These included antiparticles as well as high-energy gamma ($\gamma$) rays and neutrinos \cite{Silk:1984zy}.
 
Prior to the launch of the \textit{Fermi} satellite in 2008, the specific prediction was made that the Milky Way (MW) galactic center (GC) would be an ideal region to search for such a high energy signature on the basis of indications of an excess diffuse microwave central glow discovered by the WMAP satellite \cite{WMAP:2003cmr} and confirmed by Planck \cite{Planck:2013intres}. Early suggestions were that this excess was galactic-dust related, but this required implausible dust models \cite{Finkbeiner:2003im}. Subsequently, it was proposed that the excess could be due to synchrotron emission from relativistic electrons and positrons generated in WIMP dark matter annihilation \cite{Hooper:2007kb, Goodenough:2009, Hooper:2011a, Hooper:2011b, Daylan:2016}.
These authors predicted an extended diffuse $\gamma$-ray signal in the energy range up to a few hundred GeV that was subsequently detected by the \textit{Fermi}-LAT \cite{vitale2009indirectsearchdarkmatter} and identified with some four years of data as the Galactic Center Excess (GCE) \cite{Abazajian:2012pn}. However, it was soon recognized that the GCE morphology was possibly due to unresolved astrophysical sources while still being consistent with the DM WIMP prediction \cite{bartels2018, 2018NatAs...2..387M, 2019JCAP...09..042M}. 
Despite decades of refinement of both the WIMP theory, GCE data, and corollary searches for signals, for example, due to unresolved GC sources, similar WIMP DM signal in nearby dwarf spheroidal galaxies and in M31, and in direct deep underground WIMP searches, the so-called WIMP window \cite{Leane:2018kjk} remains open for plausible candidate channels such as $b,\bar{b}$ annihilations that fit the GCE $\gamma$-ray spectrum for WIMP masses $m_\chi\sim \qty{50}{\giga\electronvolt}$. 

The dark matter community is at a crossroads in the absence of any direct detection results from a number of deep underground experiments that are now rapidly approaching the neutrino floor. Our only possible evidence for WIMPS comes from the highly significant \textit{Fermi}-LAT detection of the GCE. At present, millisecond pulsars (MSPs) present the best (astrophysical) explanation, although doubts have been raised about the possible MSP flux as it requires a new population of weak sources \cite{Holst:2024fvb} as well as the fit to the masked data for the predicted MSP morphology \cite{Zhong:2024vyi}. 

The \textit{Fermi} data \cite{Zhou:2015, Ajello:2016, Fermi:2017} provides strong evidence for the Galactic Center Excess.
The significance of the GCE is remarkably high \cite{Calore:2015b, Di_Mauro_2021} despite the various foreground corrections.
Uncertainties include cosmic-ray source distributions and their interaction with interstellar gas in the Milky Way, a potential contribution from the Fermi bubbles, and resolved point sources of gamma rays. The spectral template advocated by the \textit{Fermi} Collaboration seems to us to generate the cleanest GCE map, and we note, in common with many previous discussions \cite{Abazajian:2010zy, Bartels:2015aea, Ramirez:2024oiw, Malyshev:2024qer, Gordon:2013vta, Manconi:2024tgh}, that a flattened asphericity is present in the central kpc region, however, for an alternative point of view see \cite{McDermott:2022zmq}. It is this asphericity that is included, and indeed derived, in the modeling of old stars (MSP) but has hitherto been ignored in all DM modeling of the GCE. We seek to rectify this situation, to provide a proper basis of comparison between the rival models. To anticipate our conclusion, we find that morphologies of the GCE predicted by old stars (MSP) and DM annihilation models are essentially indistinguishable. 

We point out here that the WIMP DM model for the GCE is incomplete. There are three potential signals that distinguish the astrophysical and particle models for the GCE. These are a) the predicted $\gamma$-ray spectrum: both models pass this test; b) the predicted $\gamma$-ray flux: both models pass this test, subject to qualifications previously cited; and c) the $\gamma$-ray morphology: the MSP model passes this test, but the DM model has not been robustly tested. Hitherto, all analyses to our knowledge of the DM explanation of the GCE have adopted 
spherically symmetric models (i.e., generalized NFW) for the DM. We show here that this is insufficient, and we provide improved models based on recently acquired knowledge of the complex MW galaxy merging history in the first \qty{3}{\giga\year}. We use numerical simulations with constrained initial conditions to elucidate the DM morphology.

\section{Simulations} \label{sec:simulations}

To study the morphology of MW-like galaxies in high-resolution, we make use of the High-resolution Environment Simulations of The Immediate Area (\textsc{Hestia}) simulations suite \cite{HESTIA2020}. We use high-resolution magnetohydrodynamical cosmological zoom-in simulations. All the \textsc{Hestia} simulations use initial conditions that are constrained using peculiar velocities from CosmicFlows-2 \cite{Tully:2013} in the framework of the Constrained Realizations technique \cite{HoffmanRibak:1991, Doumler:2013a, Doumler:2013b, Doumler:2013c, Sorce:2015}. The resulting simulations contain a Local Group in the true cosmographic environment, including Virgo and the Local Void to match the actual observed Universe. The high-resolution simulations achieve an effective resolution of $8192^3$ particles within two overlapping \qty{3.7}{\mega\parsec} spheres around the two main galaxies. The mass resolutions are $m_\mathrm{dm} = \qty{1.5e5}{\Msol}$ and $m_\mathrm{gas} = \qty{2.2e4}{\Msol}$, for DM and gas, and the softening length is $\qty{220}{\parsec}$. \textsc{Hestia} simulations assume $\Lambda$CDM cosmology in accordance with \textit{Planck} Collaboration (2014)\cite{Planck:2013}, namely $\sigma_8 = 0.83$, $H_0 = \qty[per-mode=power]{67.66}{\kilo\meter \per \second \per \mega\parsec}$, $\Omega_\Lambda = 0.682$, $\Omega_m = 0.270$, and $\Omega_b = 0.048$.

The \textsc{Hestia} simulation suite uses Amiga's Halo Finder (AHF) \cite{AHF2009}. AHF identifies halos and subhalos and their properties by detecting gravitationally bound particles using all particles, not just DM. The halo finder provides, among other things, the halo centers, masses, and shape of the inertia tensor. The AHF also includes a halo merger tree builder package.

We analyze the three \textsc{Hestia} high-resolution simulations. Each simulation hosts a Local Group of two similarly sized galaxies, of which the (slightly) less massive one is usually dubbed the MW, but in practice, both are MW-like in terms of mass, morphology, and other characteristics. Table~\ref{tab:galaxies} lists the properties of the six MW-like galaxies. To emulate observations, we calculate projected density maps around the galactic bulge. The observer is set on the galactic disc plane \qty{8}{\kilo\parsec} from the center of the galaxy. We show three different kinds of maps: a) DM density maps, which only account for dark matter in the simulation; b) quadratic DM density maps for better comparison with the annihilation cross-section; and c) stellar density maps, which only contain stars older than 3 Gyr to emulate a possible MSP population. We also calculate the minor-to-major axis ratios of the projected density maps at different density thresholds to quantify the asphericity of the morphology. For a more detailed description of projected density map calculation, see Appendix~\ref{sec:methods}.

\begin{table}
\begin{ruledtabular}
    \caption{Properties of the six MW-like galaxies in the \textsc{Hestia} high-resolution simulations. The mass within a sphere with a mean density equal to $200\rho_\mathrm{crit}$ is given in the column $M_\mathrm{200c}$, and the radius of that sphere in the column $R_\mathrm{200c}$. The stellar mass of the disc is given in the column $M_\mathrm{star,disc}$. The column $R_\mathrm{bulge}$ shows the effective radius of the Sérsic profile describing the bulge.}
    \centering
    \begin{tabular}{c c c d c d}
        ID & Simulation & $M_\mathrm{200c}$ & \multicolumn{1}{c}{\textrm{$M_\mathrm{star,disc}$}} & $R_\mathrm{200c}$ & \multicolumn{1}{c}{$R_\mathrm{bulge}$} \\
        & & \qty{e12}{\Msol} & \multicolumn{1}{c}{\textrm{\qty{e10}{\Msol}}} & \unit{\kilo\parsec} & \multicolumn{1}{c}{\unit{\kilo\parsec}} \\ \colrule
        G1.1 & 09\_18 & 1.94 & 7.70 & 263 & 2.88 \\
        G2.1 & 17\_11 & 1.96 & 8.38 & 264 & 1.95 \\
        G3.1 & 37\_11 & 1.04 & 5.70 & 212 & 0.71 \\
        G1.2 & 09\_18 & 2.13 & 11.4 & 271 & 2.11 \\
        G2.2 & 17\_11 & 2.30 & 9.90 & 279 & 2.32 \\
        G3.2 & 37\_11 & 1.09 & 4.12 & 214 & 1.99 \\
    \end{tabular}
    \label{tab:galaxies}
\end{ruledtabular}
\end{table}

\begin{figure*}
    \centering
    \includegraphics[width=0.9\textwidth]{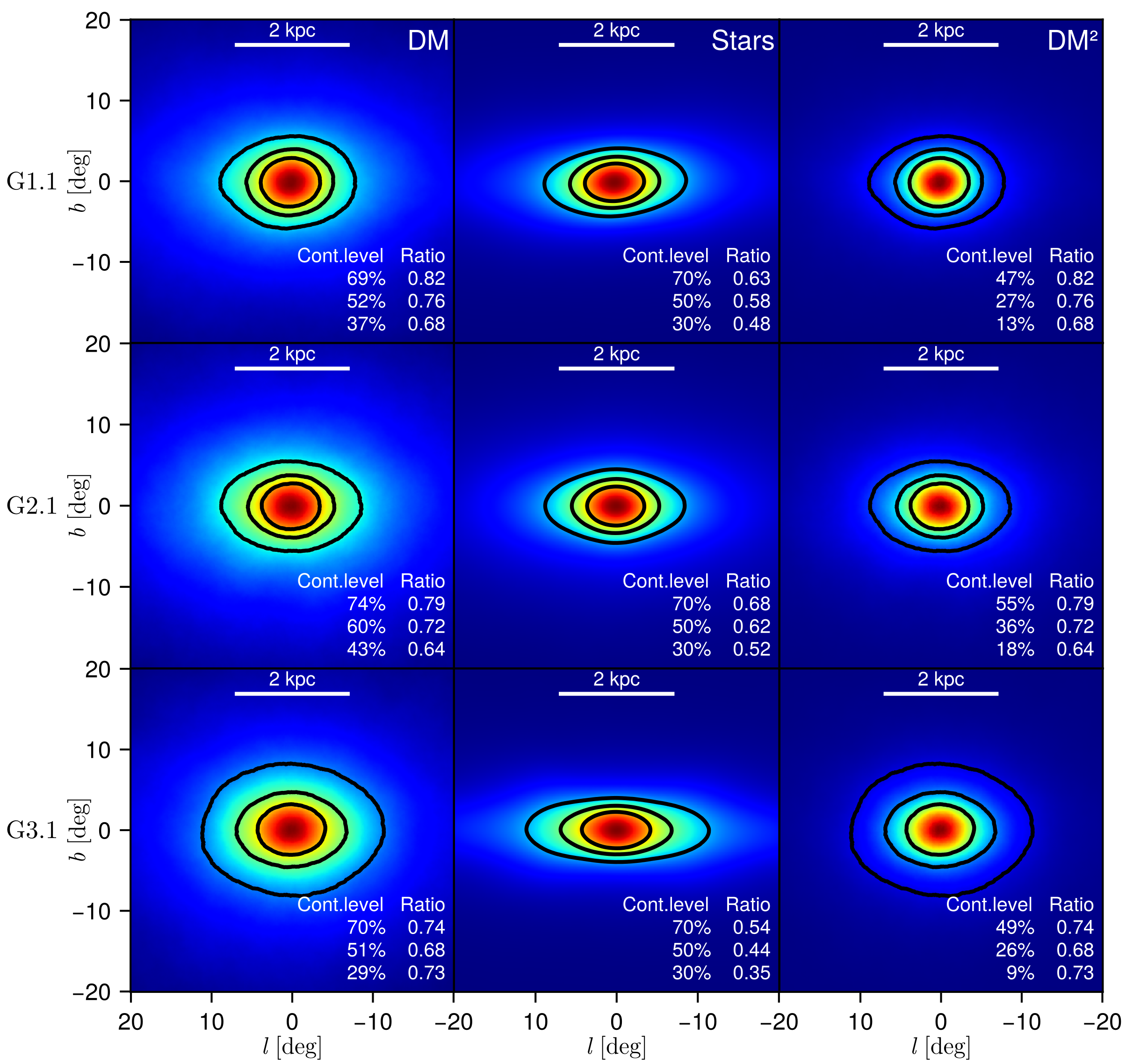}
    \caption{Normalized density projections of the bulges. The columns show dark matter, stellar, and quadratic dark matter densities, respectively. Only stars older than \qty{3}{\giga\year} are used to calculate the stellar density projections. Rows depict different galaxies. Higher densities are shown in red, and lower densities in blue in linear scale. The lower right corner of each plot lists the levels of the isodensity contours as a percentage of the maximum and axis ratios for corresponding contours. The calculation of axis ratios is described in Appendix~\ref{sec:methods}. The depicted spatial scale corresponds to the scale at the center of the galaxy. The galaxy properties are listed in Table~\ref{tab:galaxies}. This figure is continued in Fig.~\ref{fig:dm-stars-dm2-2}.}
    \label{fig:dm-stars-dm2-1}
\end{figure*}

\begin{figure*}
    \centering
    \includegraphics[width=0.9\textwidth]{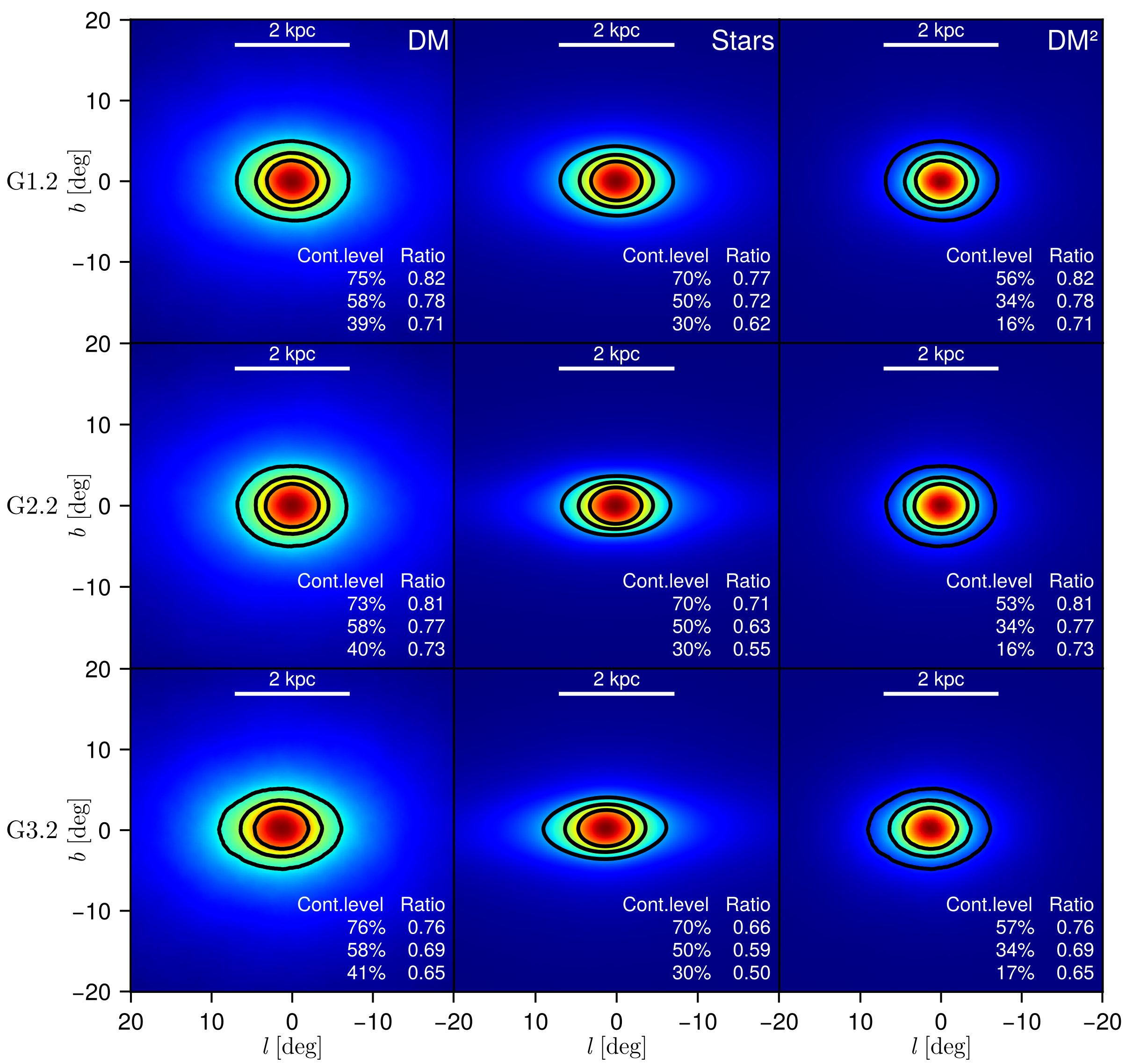}
    \caption{Normalized density projections of the bulges. Continued from Fig.~\ref{fig:dm-stars-dm2-1}}
    \label{fig:dm-stars-dm2-2}
\end{figure*}

\section{Results} \label{sec:results}

The dark matter halos around galaxies are rarely as spherical as the isotropic NFW model explicitly assumes \cite{Allgood:2006, Calore:2015, Chua:2019}. We show that this also holds for the central regions of the galaxy around the bulge. The analysis of the MW analogs in \textsc{Hestia} simulations shows a flattened DM density projection for all the galaxies. Fig.~\ref{fig:dm-stars-dm2-1} and Fig.~\ref{fig:dm-stars-dm2-2} compare the dark matter, stellar, and quadratic dark matter projected density around the galactic center for all six galaxies. The axis ratios of the isodensity contours (in the lower right corner of each plot) show increasing flatness for the outer parts of the galactic bulge. A spherical profile, such as NFW, would have an axis ratio of \num{1.0}. In the most extreme cases, the axis ratios of dark matter density projections of simulated galaxies reach about 2:3.

Stellar density projections are calculated using stars older than \qty{3}{\giga\year} to emulate a possible MSP population. As expected, these density projections exhibit flattened and boxy morphology, but can also be affected by the disc component. Both the quadratic DM and stellar density projection are similar in the vertical axis, but the stellar component is much more extended in the disc plane, which results in lower axis ratios. For completeness, we studied the redshift evolution of the density profile. Appendix~\ref{sec:redshiftevo} details the evolution of a galaxy both in DM and stellar density profiles.

In summary, our constrained simulation results for the dark matter distribution in the inner bulge demonstrate for the first time that DM annihilation maps cannot be modeled by a spherical NFW density profile. The merger history chosen to match the GAIA reconstruction of the last few billion years inevitably generates a non-axisymmetric DM distribution. This anisotropy is amplified as we compute the projected quadratic projection. The models presented here demonstrate a boxiness in the predicted $\gamma$-ray maps of the GCE that is similar to that found in the old bulge population used to generate MSP diffuse emission maps. Given the uncertainties in the \textit{Fermi}-LAT GCE data due to foreground cleaning, we infer that the predicted DM and MSP morphologies are indistinguishable.

\section{Conclusions}

The WIMP DM hypothesis is at a crossroads. The GCE detected by \textit{Fermi}-LAT is highly significant at nearly $40\sigma$. But our community is split between rival hypotheses, which are often alleged to disfavor the dark matter option. This paper addresses one of the main objections against the DM hypothesis, namely that the observed GCE morphology is most likely not spherically symmetric, as has been generally anticipated for the DM NFW profile and the associated annihilation signal. We have demonstrated that realistic DM modeling based on constrained numerical simulations of the MW galaxies indicate that the DM distribution in the inner kpc possesses a significantly flattened and asymmetrical morphology that is compatible with boxy profiles and is indistinguishable from modeling of \textit{Fermi}-LAT GCE by independent groups of both the GCE data via the WIMP annihilation model with, for example, annihilations via $\sim40$ GeV $b, \bar b$ channels, and equally for the old star population morphology as expected for the MSP distribution.

There are four properties of the GCE modeling of prime importance to test our hypothesis.
\begin{itemize}
\item {\bf Spectrum.}
With current data, the two hypotheses are indistinguishable over the 1-10 GeV range. This is expected to change in the near future, see below.
\item{\bf Normalization.}
The sweet spot for the WIMP is the lightest stable supersymmetric particle of mass  $\sim$ \qty{50}{\giga\electronvolt} that is a hadronic counterpart of the w-quark or Higgs boson with annihilation channels including quarks,  muons, and potentially observable neutrinos, and a thermal freeze-out cross-section
\cite{rott2023}. This gives the observed normalization for the \qtyrange{1}{100}{\giga\electronvolt} flux.
The MSP hypothesis has a substantial problem, however. Using known MSP populations in globular clusters, conjectured to be the source of the GC MSP excess, fails to give the observed GCE flux. The shortfall of up to an order of magnitude is only explained by hypothesizing the existence of an unobserved population of low luminosity MSPs \cite{hooper2024}. Whether there is a case for this unresolved faint population is the subject of ongoing debate \cite{Leane_2020,2020PhRvD.102b3023B, 2025PhRvD.111d3033M, Cholis:2022}. 
\item{\bf Morphology.}
The observed GCE morphology is noisy but distinguishably flattened over the central \qty{1}{\kilo\parsec}. The MSP distribution, best characterized by the Coleman bulge \cite{Coleman:2020}, is indistinguishable. We have demonstrated that the DM generates a similar boxy distribution for the GCE emission.

We note that a wide variety of opinions about GCE morphology as a probe of the rival hypotheses of DM and MSP interpretations can be found in the literature, as referenced above. However, there is one robust conclusion that we wish to emphasize. The data, both from the gamma-ray maps, a prediction of the WIMP interpretation, and the cospatial old stellar population, a proxy for the MSP hypothesis, strongly favor an aspherical, flattened morphology. This is well-established from modelling the \textit{Fermi}-LAT data with varied assumptions about foreground subtraction, and for the MSP hypothesis for various bulge models and data sets. Our work analyses a realistic morphology for the DM distribution and emphatically demonstrates its asphericity and flattening via a number of constrained simulations.

\item{\bf Spectral cut-off.}
DM self-annihilation has an exponential cut-off at approximately half the WIMP mass. Indeed, it is plausibly super-exponential for certain channels \cite {belikov2013}. However, MSPs may have a high-energy power law-like tail extending up to about 1 TeV. Current data is suggestive of such a tail, which may, however, be due to point sources rather than any diffuse excess \cite{linden2016}.
\end{itemize}

Both hypotheses for the GCE, that of DM annihilations and MSPs, are equally plausible based on morphology, spectrum, and intensity, with perhaps a slight edge for the DM hypothesis on the last of these attributes in view of the observed deficiency in MSPs. Imminent observations, both of the GCE and nearby dwarf galaxies, should play a critical role in adjudicating this tension between eminently plausible theoretical explanations. We expect that CTAO and SWGO observations will provide the ultimate constraints on the contribution of MSPs or indeed of DM to a possible high-energy tail of the GCE excess in the near future.

Other WIMP physics may be buried in the GCE.
For example, in SUSY, the annihilation cross-section that is specified by the symmetry breaking scale is of order \qty{1}{\tera\electronvolt} or larger from LHC constraints \cite{Rodd:2024qsi} on the canonical SUSY model. Such a high DM candidate particle mass requires DM to be in the form of SUSY relic winos or higgsinos in the \unit{\tera\electronvolt} mass range. The predicted self-annihilation gamma rays are $\sim \qty{0.1}{\tera\electronvolt}$ and constitute an important target for future $\gamma$-ray observatories (including CTA, CTAO-South, and SWGO).
We also note that the range of velocity dispersion dependences predicted in the various cross-sections will lead to novel morphological differences that will help explore new DM physics.

\begin{acknowledgements}
MMM was funded by the Estonian Ministry of Education and Research (grant TK202) and the Estonian Research Council grant (PUTJD1203). MMM and NIL acknowledge the funding by the European Union's Horizon Europe research and innovation program (EXCOSM, grant No. 101159513). 
YH is partially supported by the Israel Science Foundation (ISF 1450/24).
The authors gratefully acknowledge the Gauss Centre for Supercomputing e.V. (www. gauss-centre.eu) for funding this project by providing computing time on the GCS Supercomputer SuperMUC at Leibniz Supercomputing Centre (www.lrz.de).
The analysis and plotting presented here is available in \cite{gce-in-hestia-repo} and made use of Julia Language \cite{Julia:Language} and its ecosystem, specifically Makie.jl \cite{Julia:Makie} and DataFrames.jl \cite{Julia:DataFrames}.
\end{acknowledgements}

\bibliography{references}

\appendix

\section{Calculating the projected densities} \label{sec:methods}

We have produced density projections of the bulges of simulated galaxies for comparison with observational and analytical density maps by other authors. The density is calculated based on the \textsc{Hestia} simulation output, using some of the properties calculated by the AHF, as detailed below.

\begin{enumerate}
    \item From AHF halo properties, we get the unit vectors of the inertia tensor's minor and major axes. The moment of inertia tensor is calculated using particles $\leq\qty{10}{\per\h\kilo\parsec} = \qty{14.8}{\kilo\parsec}$ from the center. These unit vectors represent the orientations of the disc's normal and major axis.
    \item Using the disc's normal and major axis, we choose a position for the observer on the disc plane \qty{8}{\kilo\parsec} from the center along the major axis.
    \item Using the observer's and galactic center's coordinates, and the orientation of the normal, we calculate the galactic coordinates ($l,\,b$) for each particle.
    \item We filter the particles for the suitable particle type (dark matter or stars with chosen ages) and remove particles further than \qty{15}{\kilo\parsec} from the observer. The particles are then projected on a unit sphere using galactic coordinates.
    \item We create a regular grid of sight lines in the region of interest, i.e., $\ang{40} \times \ang{40}$ around the galactic center. Each sight line ($l,\,b$) is assigned a density value by summing up the masses of all the particles within \ang{3} of the sight line.
    \item For quadratic density, the previously calculated DM density values are squared. For stellar densities, only star particles older than \qty{3}{\giga\year} at $z=0$ are used.
    \item All the density values are normalized using the maximum density value.
\end{enumerate}

Each density map includes isodensity contour lines. For stellar maps, these contours are drawn at 70\%, 50\%, and 30\% levels of the maximum value. For DM and quadratic DM maps, the contour levels are chosen so that they intersect the $b=0$ line at the same $l$ coordinates as the corresponding stellar map contour lines. In this way, the contour lines in the stellar and DM maps are drawn approximately in the same physical region and are easier to compare.

To characterize the eccentricity of the projected density isocontours, the plots include an axis-ratio parameter for each isocontour. The lengths of the axes are estimated using Singular Value Decomposition (SVD) of the 2D shapes defined by the isocontour. The axis ratio is calculated as a ratio of the singular values of the SVD.

\section{Redshift evolution of the projected densities} \label{sec:redshiftevo}

\begin{figure*}
    \centering
    \includegraphics[width=0.8\textwidth]{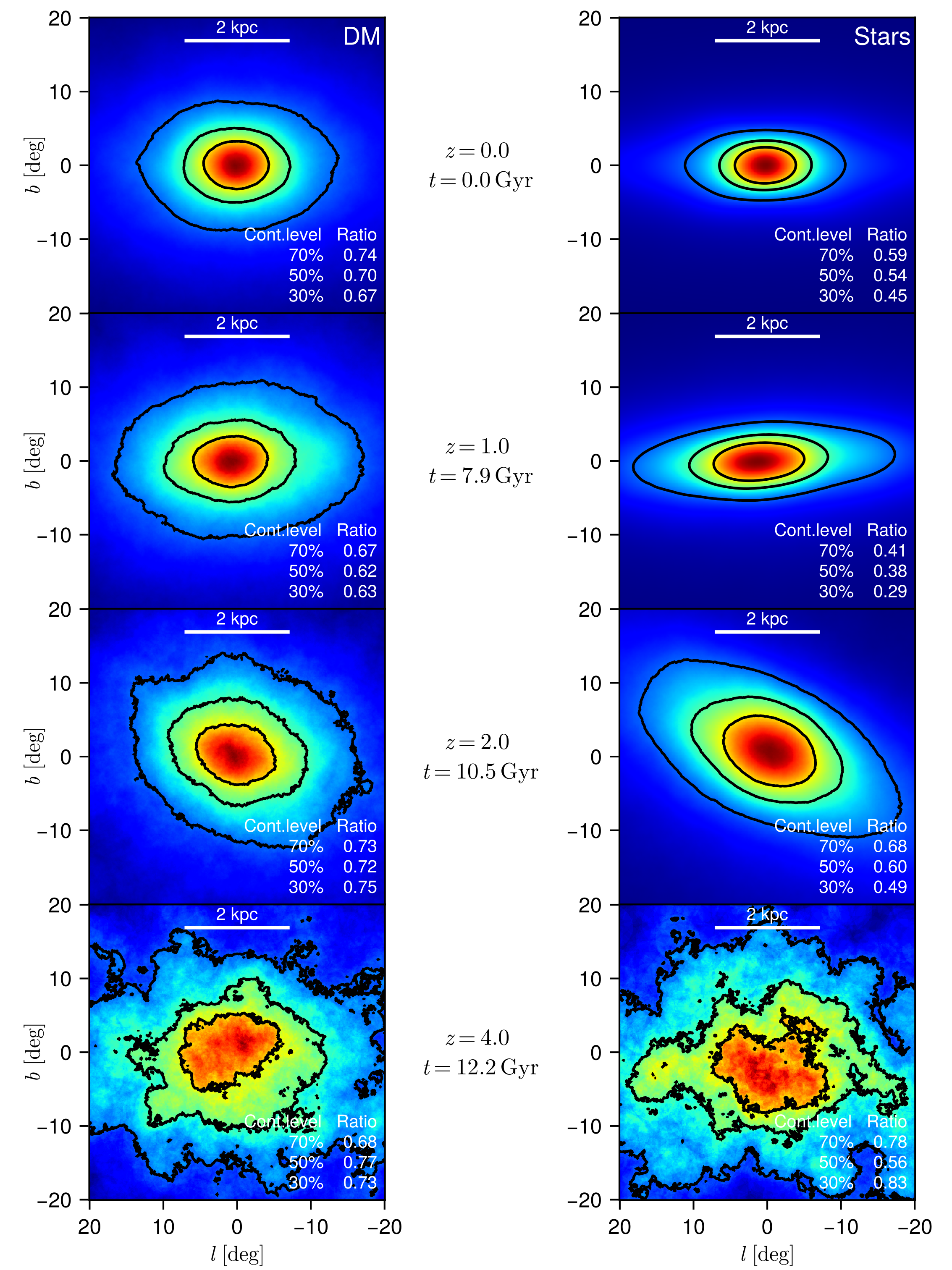}
    \caption{Redshift evolution of dark matter and stellar density projections of the bulge of G2.2. Style same as Fig.~\ref{fig:dm-stars-dm2-1}. Parameters $z$ and $t$ show redshift and lookback time.}
    \label{fig:redshiftevo}
\end{figure*}

Numerical simulations have shown that dark matter halos are not spherical and can be more accurately described by an ellipsoid. Previous studies have demonstrated, based on a set of dark matter-only simulations, how the ratio of the shortest to the longest axis depends on radius and how it evolves with redshift \cite{Allgood:2006}. In Fig.~\ref{fig:redshiftevo} we show the evolution of the shape of the dark matter and stellar distributions in the galaxy G2.2. The plotted galaxy has four significant mergers (stellar mass ratio $\geq1:10$) at lookback times \qtylist{7.28; 8.97; 9.44; 10.51}{\giga\year}. The earliest merger was the most massive one, with a stellar mass ratio of almost 1:1, and its effects are visible on the $z=2$ panel, where the bulge area appears tilted, especially so for the stellar density projection. Both the DM and stellar densities settle to their current flattened morphologies around $z=1$. As the maximum density defines the contour levels, the stellar density appears to decrease in the disk plane, which is affected by the increase of the concentration in the center. The earlier snapshots of the simulations have significantly fewer particles in the region of interest in the center of galaxies, and therefore, the density projections appear to be coarser.

\end{document}